# DOES FOREST REPLACEMENT INCREASE WATER SUPLY IN WATERSHEDS? ANALYSIS THROUGH HYDROLOGICAL SIMULATION

Ronalton Evandro Machado; Lubienska Cristina Lucas Jaquiê Ribeiro; Milena Lopes

[1]Unicamp (University of Campinas) - College of Technology. Street Paschoal Marmo 1888, CEP 13484-332, Jd. Nova Itália, Limeira, SP, Brazil.

*Correspondence to*: Ronalton Evandro Machado (machado@ft.unicamp.br)

**Abstract** – The forest plays an important role in a watershed hydrology, regulating the transfer of water within the system. The forest role in maintaining watersheds hydrological regime is still a controversial issue. Consequently, we use the Soil and Water Assessment Tool (SWAT) model to simulate scenarios of land use in a watershed. In one of these scenarios we identified, through GIS techniques, "Environmentally Sensitive Areas" (ESAs) which have watershed been degraded and we considered these areas protected by forest cover. This scenario was then compared to current usage scenario regarding watershed sediment yield and hydrological regime. The results showed a reduction in sediment yield of 54% among different scenarios, at the same time that the watershed water yield was reduced by 19.3%.

## 1. INTRODUCTION

Knowledge on how forests affect the various aspects of water is essential to assess the role of forest cover on watershed's hydrological regime (LIMA, 2010). The forest is often regarded as effective to stabilize and maintain the river flow rates and this is one of the reasons why revegetation is repeatedly recommended to recover watersheds (BACELLAR, 2005). Some of the hydrological functions usually ascribed to forests, however, such as to increase rivers water availability are disputable and lack a technical and scientific basis. We observe, however, that this is still a worldwide controversy, especially regarding the establishment of water conservation and sustainable use of natural resources policies.

In this line of research we find a large collection of data in scientific literature, resulting from watersheds systematic monitoring all over the world, which use three methodologies, especially "paired basins" (Brown, 2005). Some experiences with paired basins showed the effect of forest cover on water yield, where natural vegetation has been removed and/or replaced by planted forests (BOSCH and HEWLETT, (1982); BRUIJNZEEL (1990, 2004); BUYTAERT et al., (2006)).

The paired-basin technique would be arguably the best methodology to evaluate the hydrological functions normally assigned to forests, applicable to basins with very similar characteristics. It is always preferable that paired watersheds should be as near as possible, so as to have similar physical aspects, climate, vegetation and use and occupation (BEST et al., 2003). Despite the advantages of using paired micro-basins to study the impact of vegetation changes on water yield, this kind of study takes time, since a watershed hydrological response to tree cutting or reforestation is a medium- to long-term process. It is also impossible to test other configurations of land management and use.

Another option to predict the impact of land-use changes on the quantity and quality of water in a watershed, e.g., vegetation replacement, is the use of hydrological models. According to Sun et al. (2006) mathematical models are probably the best tools to analyze complex non-linear relationships between water yield of forests and major environmental factors.

The large number of existing models applied to watersheds shows the advancement of this technology. There are many hydrological models that simulate the quality and quantity of water flow, each one have strengths and weaknesses which must be considered according to the user's needs and the characteristics of the study area. As an example, the Soil and Water Assessment Tool (SWAT) model allows great flexibility when configuring watersheds (PETERSON & HAMLETT, 1998). The model was developed to predict the effect of different management scenarios in the quality and quantity of water, sediment yield and pollutant loads in agricultural watersheds (SRINIVASAN & ARNOLD, 1994). SWAT analyzes the watershed divided in sub-watersheds based on relief, soil and land use, preserving thus spatially distributed parameters of the entire watershed and homogeneous characteristics within the watershed.

The SWAT model is internationally recognized as a solid interdisciplinary watershed modeling tool, as demonstrated by annual international conferences and papers submitted to scientific journals (KUWAJIMA et al., 2011). SWAT many uses have shown promising results, e.g., hydrological assessments, impacts of climate change, evaluation of best management practices, estimation of pollutant load, determining the effects of land-use change, sediment yield, etc (SRINIVASAN & ARNOLD, 1994; ROSENTHAL et. al., 1995; CHO et al., 1995; MACHADO & VETTORAZZI, 2003; MACHADO et. al. 2003; KOCH et al., 2012; LESSA et al., 2014; ABBASPOUR et al., 2015; DECHMI & SKHIRI, 2013; LIU et al., 2015; ZHANG et al., 2014; ROCHA et al., 2015; LIN et al., 2015).

Due to the uncertainty of forests role in rivers produced water quantity and quality and the possibility of creating different scenarios that are difficult to test at watershed level, this paper's objective was first to identify "Environmentally Sensitive Areas" (ESAs) in the watershed under study and, subsequently, to simulate land use scenarios comparing them as to sediment yield and hydrological regime.

## 2. METHODOLOGY

**2.1. Area of study** Pinhal River watershed is located between UTM coordinates 250,000 and 275,000 m (S), 7,490,000 and 7,520,000 m (N) (UTM Zone 23 S, central meridian 45° W). It consists of approximately 300 Km$^2$ (Figure 1). It has a tropical highland climate – Cwa, according to Köeppen classification, with hot and humid summer and cold and dry winter, and average annual temperature of 25 °C. Average annual precipitation is approximately 1,240 mm.

The culture of sugarcane occupies most of the watershed area (42.3%), whereas the culture of citrus fruits occupies approximately 30% of the area. Much of the original forest vegetation has been destroyed in the process of land use and occupation, now scattered alongside watercourse banks (9%). The built-up area occupies 6.7%, located at the western side of Pinhal River watershed. The predominant soils in Pinhal River watershed are oxisols (72%) and cambisols (19%).

The Pinhal River is important for being the source of water for Limeira, state of São Paulo. The watershed has suffered in the past few decades from environmental degradation. The current situation may compromise this water source, if the process of degradation continues.

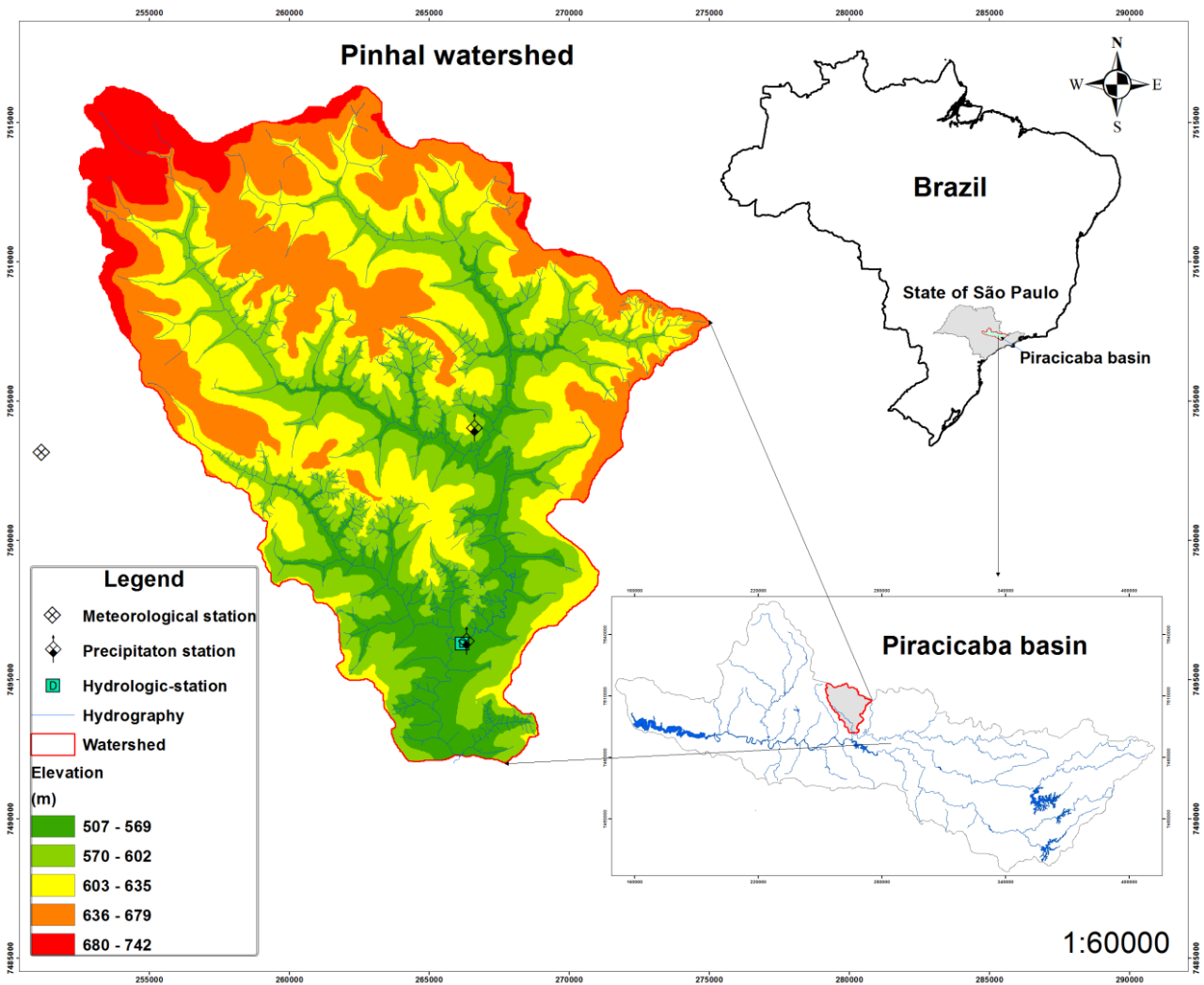

Figure 1 – Locations of the Pinhal watershed and gauging stations.

## 2.2. The SWAT model and input data

SWAT is a distributed parameter model which simulates different physical processes in watersheds and which aims at analyzing land-use changes impacts on surface and subsurface runoff, sediment yield and water quality in agricultural watersheds that were not instrumented (SRINIVASAN & ARNOLD, 1994). The model operates on a daily basis and can simulate periods of 100 years or longer to determine the effects of management changes. It has been widely applied in hydrological modelling, water resources management and water pollution issues (DOUGLAS et al., 2010).

SWAT uses a command structure to propagate runoff, sediments and agrochemicals across the watershed. The model components include hydrology, climate, sediments, soil temperature, crop growth, nutrient and pesticide loading, and agricultural management (ARNOLD et al., 1998). The hydrological component of SWAT includes subroutines of surface runoff, percolation, lateral subsurface flow, return flow of shallow aquifer and evapotranspiration.

SWAT uses a modified formulation of the Curve Number (CN) method (USDA-SCS, 1972) to calculate surface runoff. The Curve Number method relates runoff to soil type, land use and management practices (ARNOLD et al., 1995). Sediment yield is estimated using the Modified Universal Soil Loss Equation (MUSLE) (WILLIAMS & BERNDT, 1977).

The model requires as input data daily precipitation, maximum and minimum air temperatures, solar radiation, wind speed and relative humidity. Data were obtained from UNICAMP Faculty of Technology weather station, located in Limeira, state of São Paulo, at UTM coordinates 251145 m (W) and 7503161 (S). Rainfall data were obtained from two other rainfall stations (Figure 1). Other data are cartographic layers: Digital Terrain Model (DTM), Land and Soil Use. Soil physical and hydraulic properties and crop phenological properties are stored in the model database. Table 1 summarizes the input data used in the study. Inputting data (layers and alphanumeric data) into SWAT is made via appropriate interface. The interface (ARNOLD et al., 2012) was developed between SWAT and GIS ArcGis. The interface automatically divides the watershed in sub-watersheds from DTM and then extracts input data from layers and Geodatabase for each sub-watershed. The interface display the model outputs using ArcGis charts and tables. We divided the Pinhal River watershed in 25 sub-watersheds up to the runoff measuring station at UTM coordinates 266175 m (W) and 7496308 (S) (Figure 1).

Table 1. Data sources for Pinhal Catchment.

| Input data | Data description | scale | Data sources |
|---|---|---|---|
| Land use | Land-use classification - agricultural land, forest, pasture, urban and water | 25,000 | Coordenadoria de Planejamento Ambiental, Instituto Geológico, Secretaria do Meio Ambiente do Estado De São Paulo, 2013 |
| Soil | Soil types and physical properties | 100,000 | Instituto Agronômico de Campinas |
| Topography | Digital Elevation Map (DEM) | 10,000 | Instituto Geográfico e Cartográfico São Paulo |
| Hydrological and Meteorological | precipitation, minimum and maximum temperature, solar radiation, wind speed | Daily | ANA, FT |

### 2.3. Model evaluation

During analysis period (2012 to 2014) calibration of model is not possible due to inconsistency in observed data (the measuring station was constantly drowned during the operating period of a reservoir associated with a power station).

Despite the impossibility of calibrating the model for the Pinhal hydrographic basin, we used the hydrological regionalization methodology to validate the behavior of the model (Vandewiele, 1995; Bardossy, 2007). A hydrological regionalization is a technique that allows to transfer information between watersheds with similar characteristics in order to calculate, in sites where there are no data on the hydrological variables on interest (Emam et al., 2016). This technique becomes a useful tool for water resource management, especially when applied to most important instruments of Brazilian water resource policy that are the concession of water resource use rights and charging for the use of water resources (Fukunaga et al., 2015).

According to Tucci (2005), the hydrological information that can be regionalized can be in the form of variable, parameter or function. The hydrological function represents the relationship between a hydrological variable and one or more explanatory or statistical variables, such as flow-duration curve or relationship between impermeable areas and housing density (Tucci, 2002). The flow-duration curve relates the flow or level of a river and the probability of flowing greater than or equal to the ordinate value, thus being a simple, but concise and widely used method to illustrate the pattern of flow variation over time (Naghettini and Pinto, 2007).

For the construction of the flow-duration curve in this work, the series of simulated flows in the period from 2012 to 2014 was initially ordered decreasingly. This series was statistically divided into 10 equal intervals. For each interval, the number of flows was counted and the respective cumulative frequencies of the interval from highest to lowest are calculated. For the purposes of comparison, in the same graph, we plot the regionalised flows, according to the State Department of Water and Electric Energy (DAEE - state entity responsible for granting concessions of water resources in the State of São Paulo) and simulated, allowing The verification of sub or overestimation by the simulated curve. The Nash-Sutcliffe model efficiency coefficient (Nash and Sutcliffe, 1970) was used to validate the simulation results, besides the visual analysis of simulated flow-duration curve regionalized (NSE). The NSE (equation 2) was used to compare the regionalized and simulated flows in intervals of 5 in 5% probability of occurrence of the flow-duration curve. NSE can range from -∞ to 1, where 1 is the optimal value and values above 0.75 can be considered very good (Moriasi et al, 2007). NSE is calculated as eq. 2:

$$NSE = 1 - \frac{\sum_{i=1}^{n}(Q_{OBSi}-Q_{SIMi})^2}{\sum_{i=1}^{n}(Q_{OBSi}-\overline{Q}_{OBS})^2} \qquad (2)$$

The PBIAS (Eq. 3) of the simulated discharge in relation to that regionalised were too utilized (Gupta et al., 1999).

$$PBIAS[\%] = \left(\frac{\sum_{i=1}^{n}(Q_{OBSi}-Q_{SIMi})}{\sum_{i=1}^{n}(Q_{OBSi})}\right) * 100 \qquad (3)$$

Where, $Q_{OBSi}$ and $Q_{SIMi}$ corresponds to the observed and simulated discharge, respectively, on day i (m$^3$/s), and $\overline{Q}_{OBS}$ corresponds to the average observed discharge, in (m$^3$/s), and n corresponds to the number of events.

## 2.4. Identification of Environmentally Sensitive Areas (ESAs)

The concept of "Environmentally Sensitive Areas" was created in industrialized countries approximately 30 years ago due to increased soil and water degradation and the degree of severity of degradation (RUBIO, 1995). Degradation has been caused by uncontrolled forest destruction, water pollution, wind and water erosion, salinization and inappropriate management of cultivated and uncultivated soil (GOURLAY, 1998).

Environmentally Sensitive Areas (ESAs) are areas that contain natural or cultural features important for an ecosystem functioning. They may be negatively impacted by human activities and are vital to long-term maintenance of biological diversity, soil, water, or other natural resource, in the local or regional context (NDUBISI et al., 1995). An environmentally sensitive area may also be considered, in general, a specific and delimited entity with unbalanced environmental and socioeconomic factors, or not sustainable for that particular environment (GOURLAY, 1998). As an example, high sensitivity may be related to land use, which in certain cases causes soil degradation. Annual crops in areas where the relief is hilly, with declivity and shallow soils, have a high risk of degradation.

To identify ESA's in Pinhal River watershed in the context of environmental degradation, we adapted the results from Adami et al. (2012) and identified three types of ESAs: Critical, Fragile and Potential. Adami et al. (2012) made an environmental analysis of Pinhal watershed via Geographic Information System (GIS) using key indicators of relief, soil and land uses to determine the capacity of natural resources and environmental fragility. The empirical analysis of environmental fragility methodology was used to identify areas that require more attention for improving environmental conditions. The procedures employed by authors in your study are shown in Fig.2.

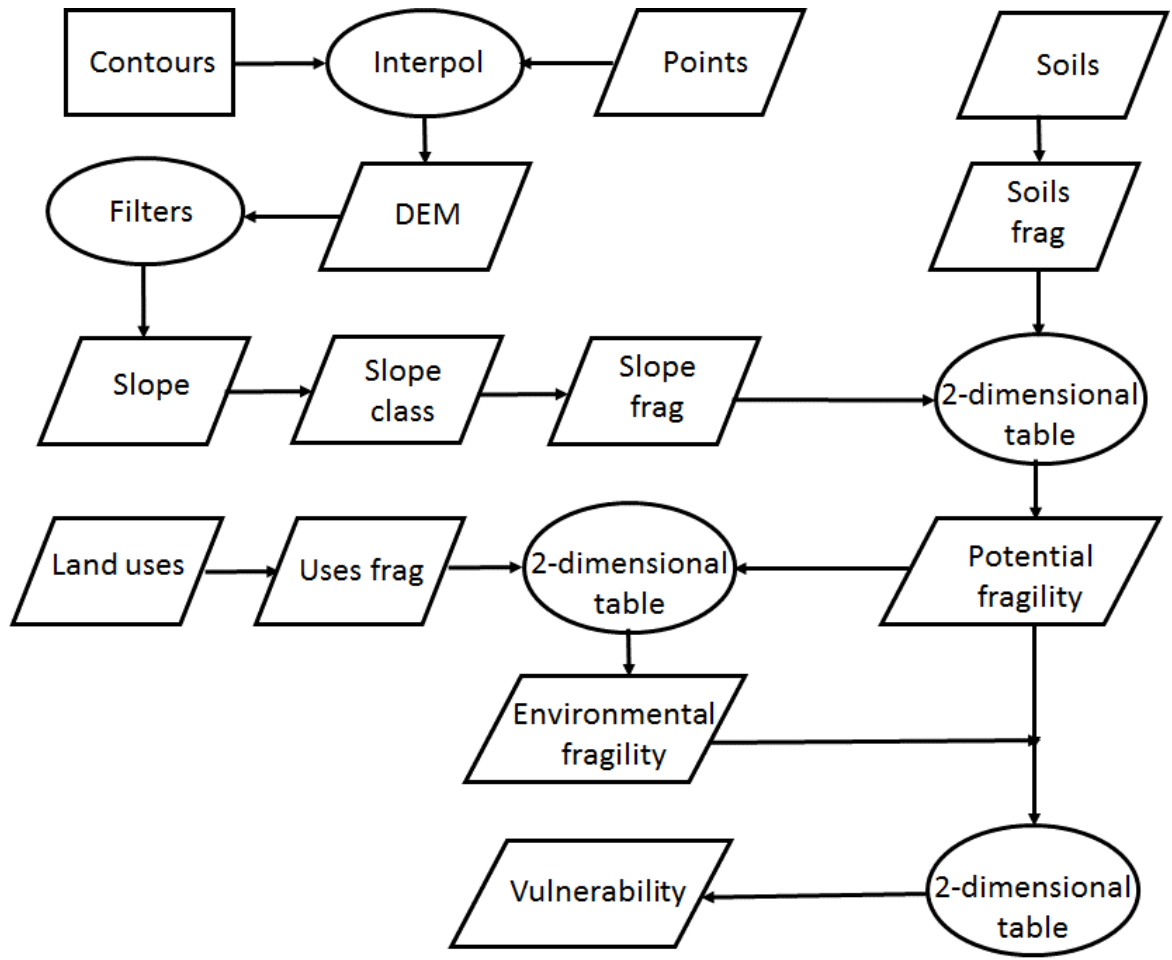

Figure 2. Flowchart of the procedures in the study the Adami et al. (2012).

## 2.5. Scenario simulation

We made two scenario simulations using SWAT model interfaced with GIS ArcGis, aiming to verify the effect of land-use change on sediment yield (sediment transported from sub-watersheds to the main channel over time, ton/ha) and the watershed hydrological regime (Discharge (m$^3$/s), surface runoff (mm), evapotranspiration (mm), soil water content (mm), water yield (mm)). Where the Water yield (mm H2O) is the net amount of water that leaves the sub-basin and contributes to streamflow in the reach during the time step. (WYLD = SURQ + LATQ + GWQ – TLOSS – pond abstractions). SURQ is the surface runoff contribution to streamflow during time step (mm H2O). LATQ is the Lateral flow contribution to streamflow during time step (mm H2O). GWQ is the Groundwater contribution to streamflow (mm). Water from the shallow aquifer that returns to the reach during the time step. TLOSS is the Average daily rate of water loss from reach by transmission through the streambed during time step (m$^3$/s) (ARNOLD et al., 2012).

One of the scenario simulations covered Critical and Fragile ESAs with overlapping forest cover on the land use map and we compared the results to the current scenario conditions (baseline). Thus, land use pattern projected in this scenario is just hypothetical and often hard to implement in practice due to already consolidated land use and occupation, but at the same time, it shows the watershed environmental fragility identified by Adami et al. (2012). Thus, these simulations illustrate the application and integration of hydrological and water quality models with GIS to evaluate watershed management scenarios modifying only land use and occupation layer and management practices.

We used the deviation of the analyzed event (PBIAS) as statistical criterion to evaluate sediment yield and compare the hydrological behavior of the watershed in different scenarios:

$$PBIAS[\%] = \left(\frac{E-E^*}{E}\right) * 100 \qquad (1)$$

where E represents baseline scenario events (current use) in the period and E* the results of the alternative scenario (ESAs) in the period. Percent bias calculation of the analyzed event (PBIAS) is important because it takes into account potential error among compared data. For this method, the higher the value of PBIAS (+ or -), the greater the difference in sediment yield and changes in hydrological regime among scenarios.

## 3. RESULTS AND DISCUSSION

### 3.1. Model evaluation

FIG. 3 show on the discharge data obtained by regionalization and simulated (i.e., flow-duration curve). The flow-duration curves generated show that the simulation tends to underestimate the discharge almost uniformly, presenting greater differences in the probabilities of 20 to 100%, and overestimating only those with lower probabilities (10 to 20%). Although underestimating the flows most of the time, the simulated flow-duration curve presented a pattern of variation close to the pattern of variation of the regionalized flows. The NSE applied to compare the regionalized and simulated flows in intervals of 5 in 5% of the flow-duration curve was 0.93. According to Moriasi et al (2007), NSE values between 0.7 and 1 indicate a very good performance of the model. As for the PBIAS result for the flow values at intervals of 5% of probability of occurrence, the model underestimated the flows by 11%. PBIAS between 10% and 15% indicates a good accuracy of the model (Van Liew et al., 2007). Emam et al. (2016), used SWAT model in the ungauged basin in Central Vietnam. The hydrological regionalization (i.e., ratio method) approach was used to predict the river discharge at the outlet of the basin. The model was calibrated with Nash-Sutcliff and $R^2$ coefficients greater than 0.7 in time scales daily by river discharge.

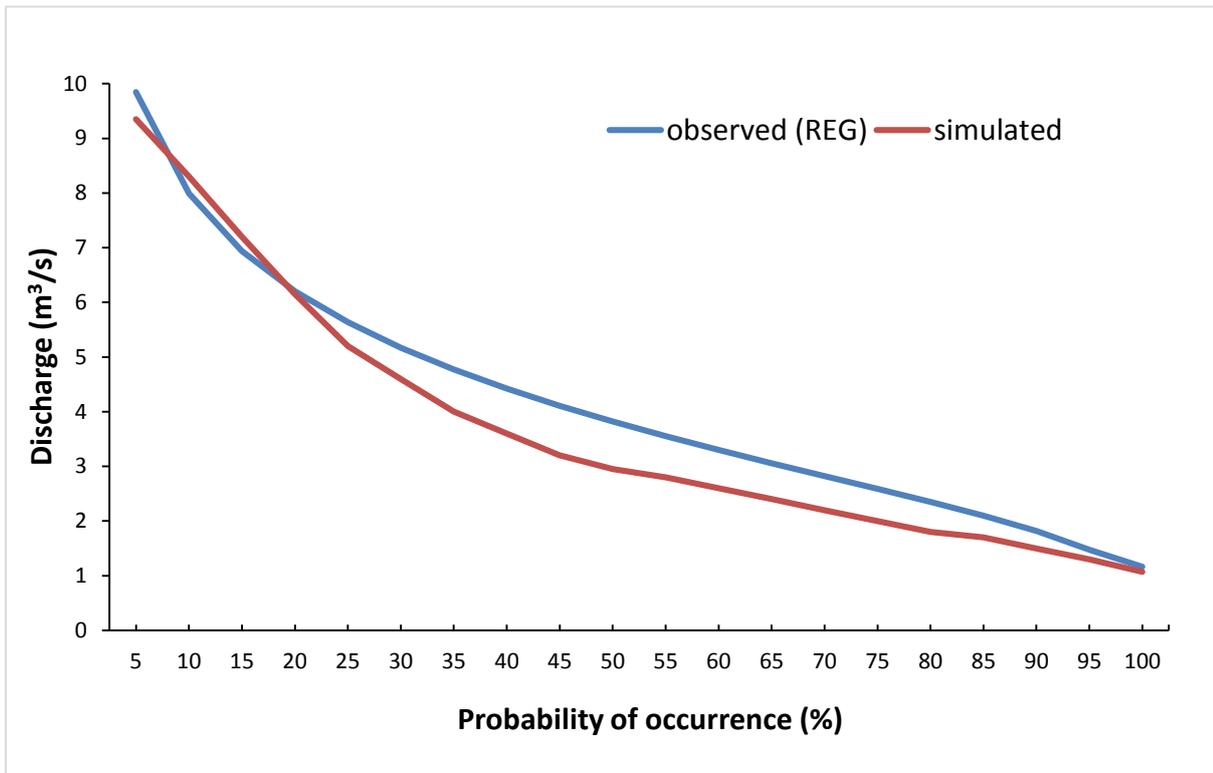

Figure 3 – Comparison of observed (hydrological regionalization) flow-duration curve with simulated on Pinhal watershed in the 2012-2014 period.

### 3.2. Environmentally Sensitive Areas (ESAs)

ESAs identified in the Pinhal River watershed are shown in Figure 4 and Table 2. 16% of the watershed area are degraded due to improper land use, which is a threat to the surrounding environment. These areas are severely eroded and have high rates of surface runoff and soil loss. In this case, there may be higher peak streamflow and water bodies sedimentation (critical ESAs).

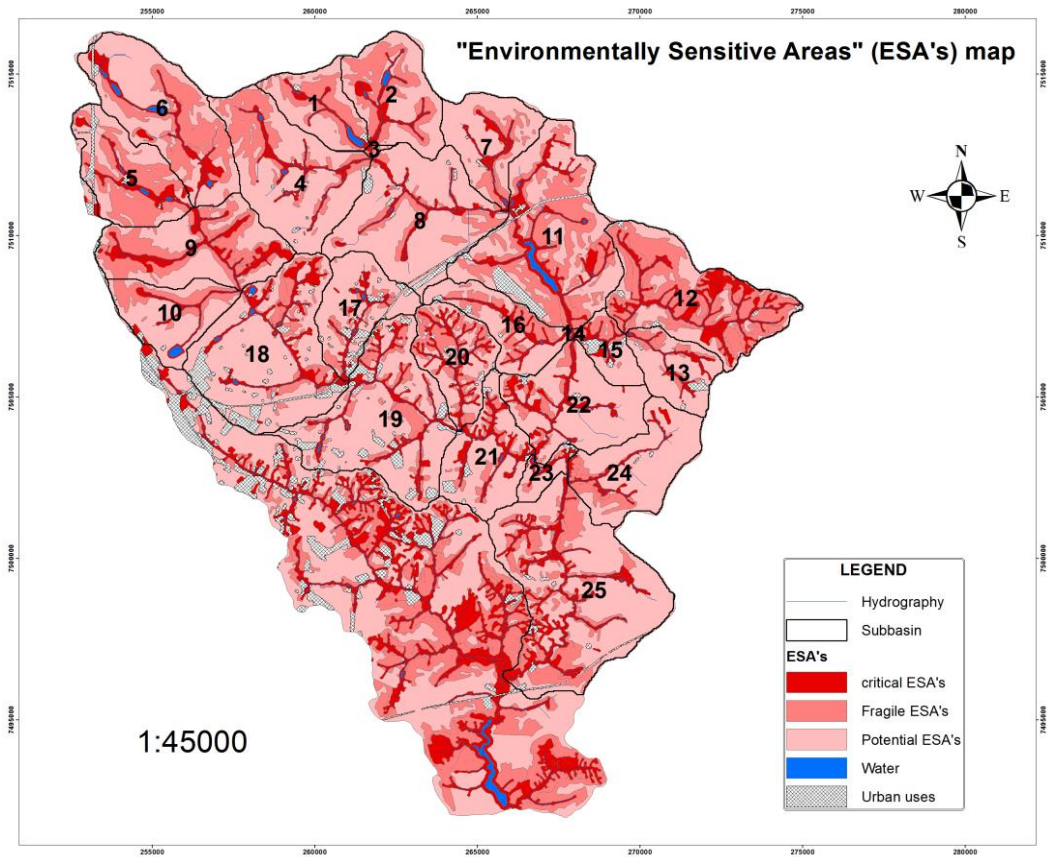

Figure 4 – ESAs Map in Pinhal River watershed.

Table 2. Environmentally Sensitive Areas (ESAs) identified in Pinhal River watershed.

| Class | Area (ha) | Area (%) |
|---|---|---|
| Critical ESAs | 4,801 | 16 |
| Fragile ESAs | 7,471 | 25 |
| Potential ESAs | 16,155 | 54 |
| Water | 149 | 1 |
| Urban or rural uses | 1,196 | 4 |
| Total | 29,772 | 100 |

In 25% of the area we have identified regions where any change in the delicate balance between the environment and human activities may result in environmental degradation of the ecosystem. A change in soil management of annual and semiannual plants, e.g., sugarcane, in highly sensitive soils may cause an immediate increase in surface runoff and water erosion, pushing pesticides and fertilizers downstream (Fragile ESAs).

54% of the total watershed area is classified as Potential ESAs. Agricultural activities in these areas, although following Land Use Capability standards and requiring simple soil conservation practices to control erosion, require attention because of the use of external agents such as pesticides in cultures of sugarcane and citrus fruits.

### 3.3. Land-use change between scenarios

Figure 5 presents the land use map for the two scenarios and Table 3 the total and relative areas of occupation of each land cover in Pinhal River watershed for current use scenario (baseline) and for the scenario of ESAs recomposed with native vegetation. From the current scenario to ESAs scenario there is a reduction of areas occupied with sugarcane, citrus and pasture and, consequently, an increase of areas occupied with forest vegetation. Sugarcane occupied the largest area in the watershed and in the ESAs scenario there was a reduction of 46.30% in this area. Orange occupies the second largest area in current use scenario and in the new scenario was reduced by 18.8%, whereas pasture was reduced by 44.43%. Other uses area has been reduced by 42.61%. Some sub-watersheds increased forest cover compared to others: sub-watersheds number 11, 12, 13, 14, 15 and 16.

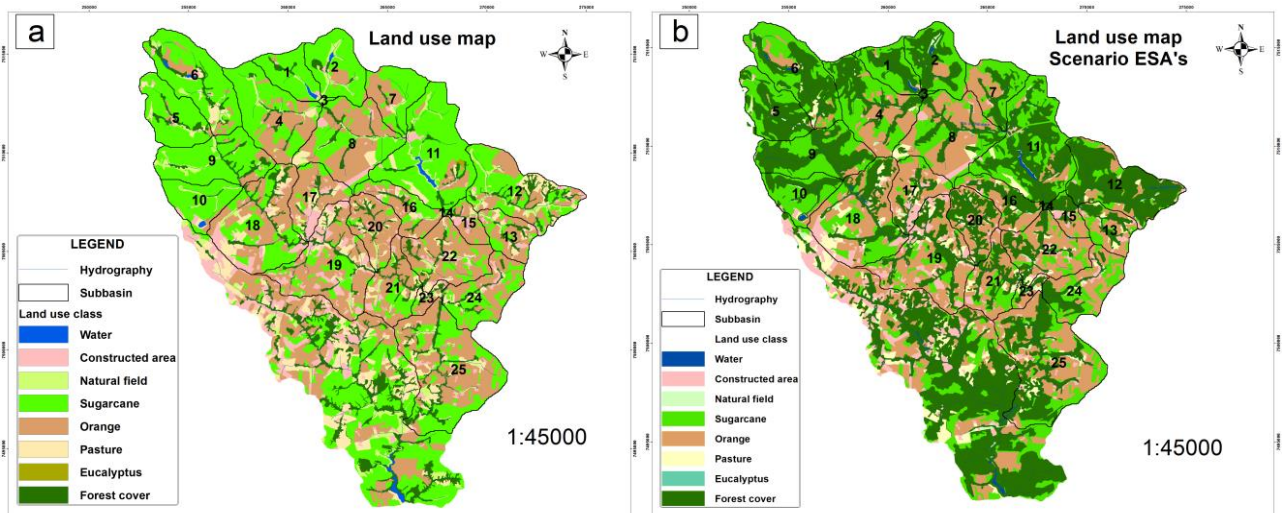

Figure 5 – Land use map for current scenario (a) (Source: Secretaria do Meio Ambiente do Estado de São Paulo, 2013) and Critical and Fragile ESAs scenario (b), with native forest cover, overlapping current land use on Pinhal River watershed (ESAs scenario).

We present in Figure 6 the variation of land-use change in sub-watersheds scale between the two scenarios. The decrease in pasture and sugarcane areas, where soils are exposed to erosion during soil management and the increase of native vegetation area explain lower sediment yield and water yield. The decrease of pasture and increase of forest area in the Northwest region (Sub-watershed 12) also contributed to lower sediment and water yield in this region.

Table 3. Land use and occupation change between the two scenarios (current use and ESAs) in Pinhal River watershed.

| Land-use type | Current use | | ESAs scenario | | Change | |
|---|---|---|---|---|---|---|
| | Area (ha) | Percentage (%) | Area (ha) | Percentage (%) | Area (ha) | Percentage (%) |
| Sugarcane | 12,566 | 42.2 | 6,748 | 22.7 | -5,818 | -46.30 |
| Orange | 8,866 | 29.8 | 7,199 | 24.2 | -1,667 | -18.80 |
| Pasture | 2,341 | 7.9 | 1,301 | 4.4 | -1,040 | -44.43 |
| Forest | 2,662 | 8.9 | 12,609 | 42.4 | 9,947 | 373.67 |
| Other uses | 3,337 | 11.2 | 1,915 | 6.4 | -1,422 | -42.61 |

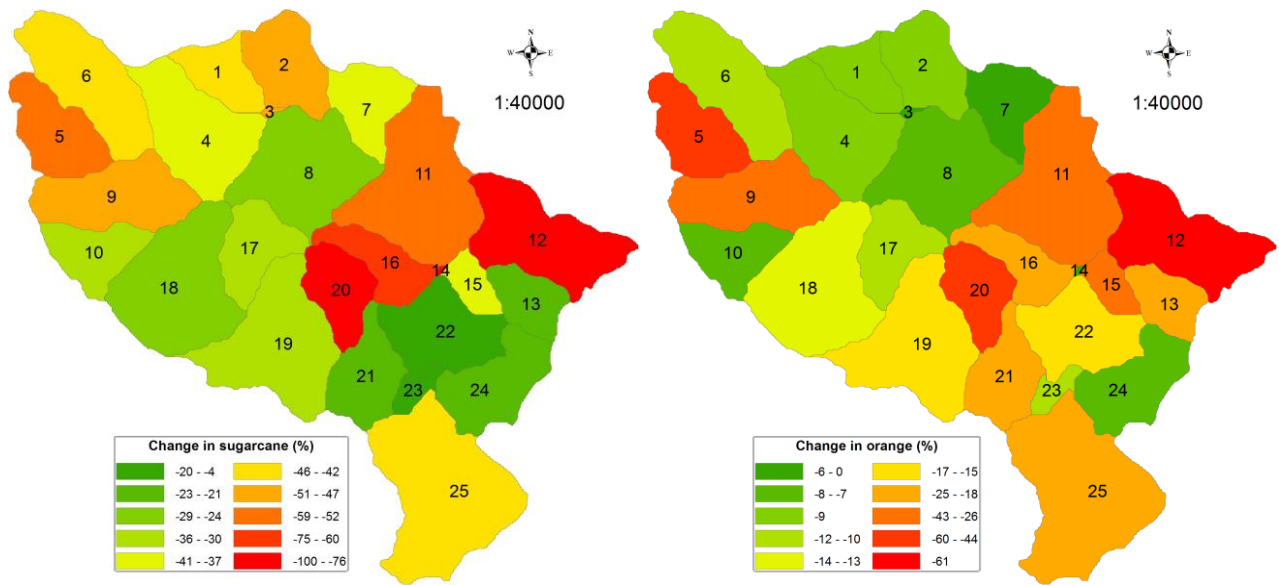

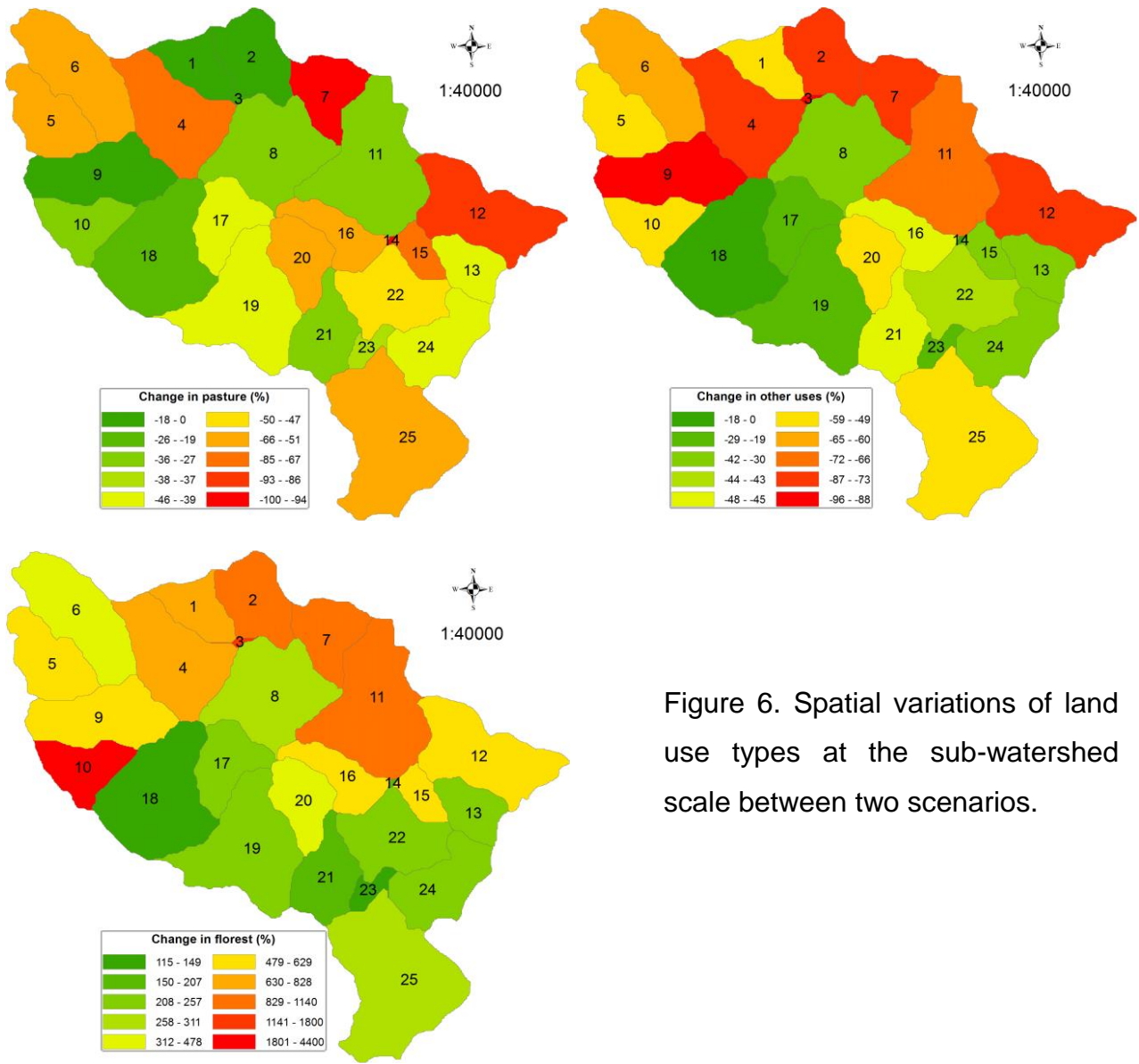

Figure 6. Spatial variations of land use types at the sub-watershed scale between two scenarios.

## 3.4. Sediment Yield

The results of sediment yield presented in Figure 7 represent the erosion and sedimentation processes occurring throughout Pinhal River watershed during simulation period (2012 to 2014). With scenario change, reduction in sediment yield was -54% (PBIAS) compared to current use scenario. This reduction occurred mostly in sub-watersheds located in lithosols and cambisols (Figure 8). These are shallow, not deep soils. Cambisols in the watershed area occurs in undulated relief. These are poorly developed soils, with incipient B horizon. One of cambisols main features is their shallowness and often high content of gravel. High silt content and low depth are responsible for this soil low permeability (TERAMOTO, 1995). The biggest issue, however, is soil erosion risk. Cambisols have restrictions of agricultural use, for their high erodibility, high risk of degradation and poor trafficability. These soils occupy 19% of the watershed total area. In current use scenario, 22.4% of this soil area is being occupied with native vegetation. In ESAs scenario the percentage increased to 68.3% (Table 4). Lithosols occupy approximately 4% of the watershed total area and are located in areas of greater declivity. They are in a geomorphologically unstable zone and erosion affects soil development and they are constantly renewed through superficial erosion (TERAMOTO, 1995). Extensive areas are occupied with sugarcane, pasture and orange (62.3%) cultivation on these soils. In the current scenario, 24.3% of lithosol is covered with native vegetation. In ESAs scenario the percentage is 95.7% (Table 4). Increased native vegetation on both soils explains 54% reduction (PBIAS) in sediment yield in the watershed, when we compare the two scenarios. Spatial location of agricultural areas in relation to relief, soil and climate is important to control erosion in watersheds.

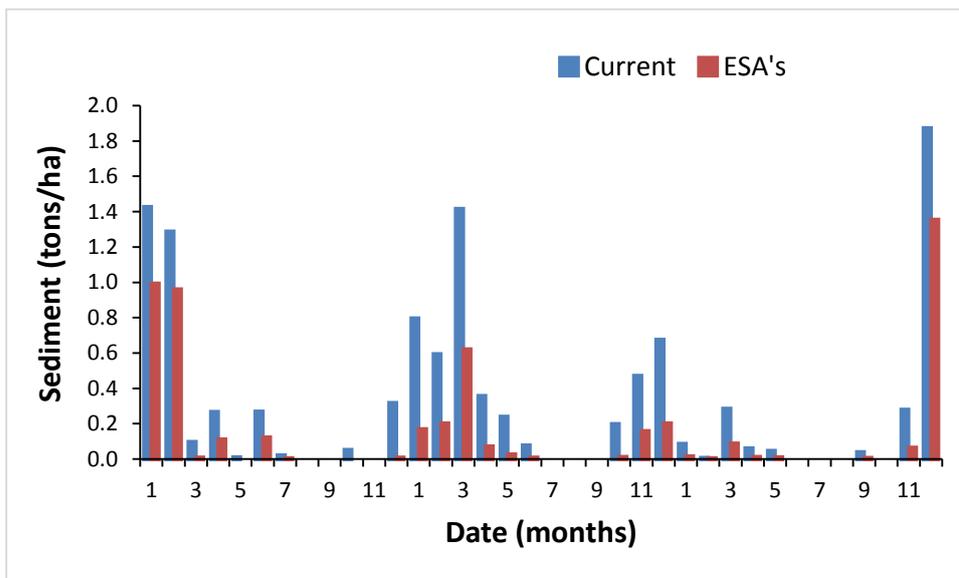

Figure 7 – Sediment yield comparison between the two scenarios on Pinhal River watershed in the 2012-2014 period.

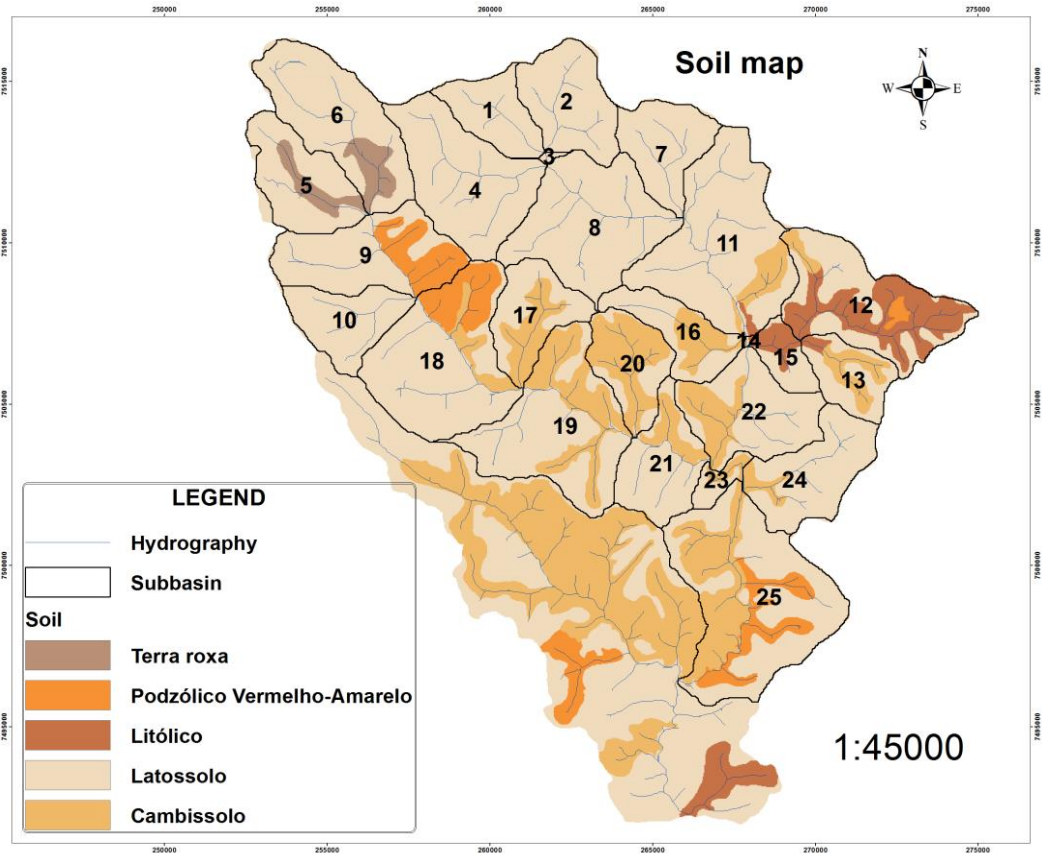

Figure 8 – Pinhal River watershed soil map (Oliveira, 1999).

Table 4. Cross tab between land-use changes on the scenarios for cambisols and lithosols in Pinhal River watershed.

| | Cambisol | | | | Lithosol | | | |
|---|---|---|---|---|---|---|---|---|
| Land-use type | Current use | | ESAs scenario | | Current use | | ESAs scenario | |
| | Area (ha) | Percentage (%) | Area (ha) | Percentage (%) | Area (ha) | Percentage (%) | Area (ha) | Percentage (%) |
| **Forest** | 1,278 | 22 | 3,894 | 68 | 275 | 24 | 1,089 | 96 |
| **Pasture** | 947 | 17 | 399 | 7 | 169 | 15 | 10 | 1 |
| **Sugarcane** | 997 | 17 | 142 | 2 | 350 | 31 | 8 | 1 |
| **Other uses** | 2,476 | 43 | 1,263 | 22 | 339 | 30 | 31 | 3 |

Spatially analysis of sediment yield in 25 sub-watersheds identified in Pinhal River watershed modeling (Figure 9) in current use scenario showed a maximum sediment yield of 80.2 t/ha, with an average of 14.6 t/ha. Maximum sediment yield occurred in upper Pinhal watershed, a more degraded area, whereas in sub-watersheds in lower Pinhal River watershed occurs aggradation, with lower sediment yield values. In ESAs scenario, replacement with native vegetation in Environmentally Sensitive Areas lead to an average sediment yield of 5.2 t/ha per year, with maximum of 14.2 t/ha. Average soil loss in sub-watersheds was near tolerable soil loss rates, which according to Leinz & Leonardos (1977) is 7.9 ton/ha for podzol and 4.2 tons/ha for lithosol. According to Figure 7, the lowest rates of sediment yield occurred in sub-watersheds with greater forest cover. As SWAT model simulates many processes in the watershed, some parameters may affect several processes (ARNOLD et al., 2012). With reduction of surface runoff in -45.8% (PBIAS) among scenarios (Table 5) due to greater soil protection, sediment yield has also been directly affected. Sediment yield difference between the two scenarios is presented in Figure 10. Analyzing Figure 10, this difference is greater in upstream sub-watersheds and in those with greater forest cover (sub-watersheds 11, 14, 15 and 16), according to Figure 5b.

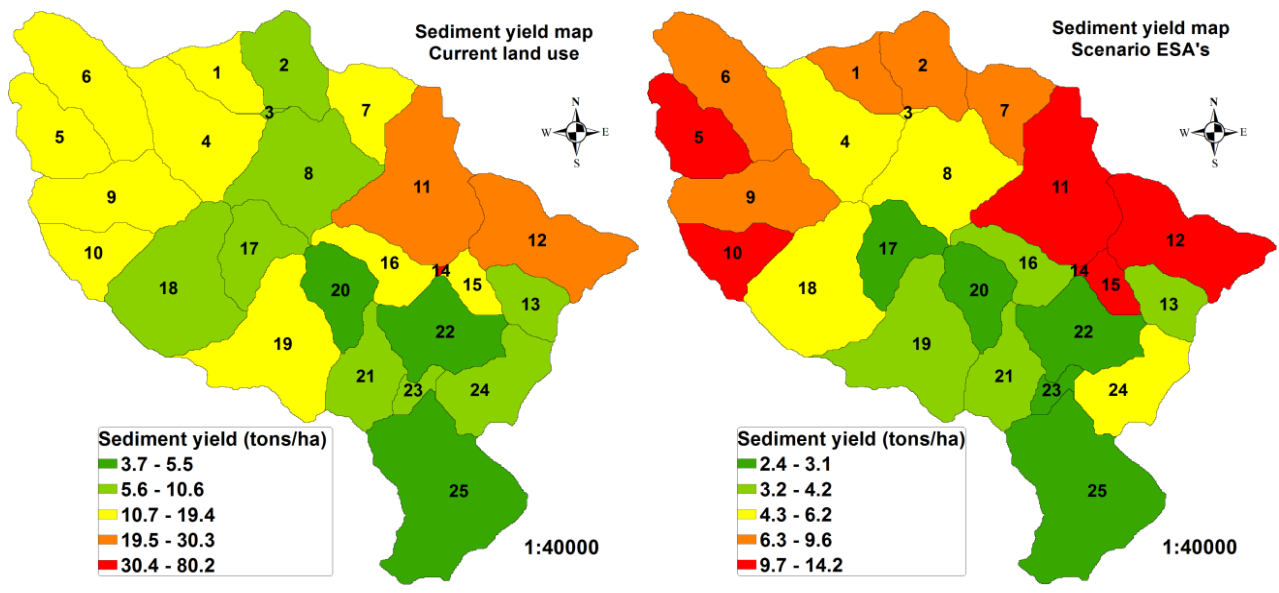

Figure 9 – Spatial distribution of average annual sediment yield at the sub-watershed scale for two scenarios.

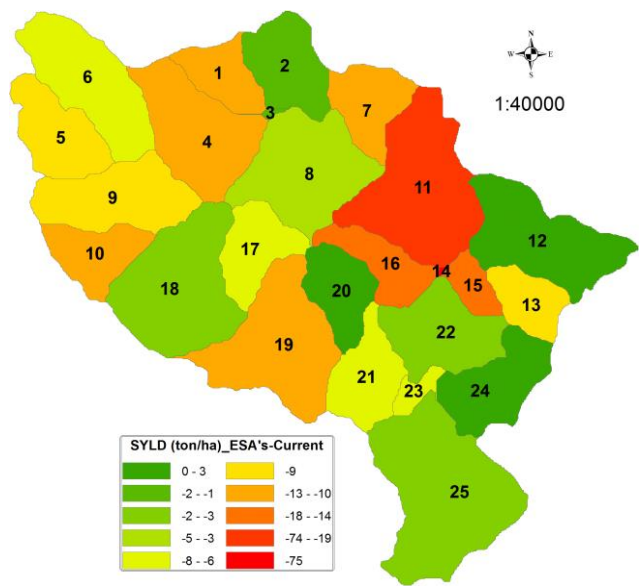

Figure 10 – Spatial variations of average annual sediment yield at the watershed scale between the two scenarios.

## 3.5. Hydrological regime

It is widely reported that land-use and land cover changes can affect quantity and quality of water resources of a watershed. We analyzed discharge ($m^3/s$), surface runoff (mm), water yield (mm), evapotranspiration (mm) and soil water content (mm) (Figures 11-15) data to evaluate the impact of these changes on the watershed hydrological regime. Monthly values for the 2012-2014 period were then compared between the two scenarios and the results (PBIAS) showed that increasing forest cover in the watershed (+ 373.8%) decreased discharge, surface runoff (SR), Soil water content (SW), water yield (WY) and increased evapotranspiration (ET) (Table 5). Studies conducted by Huang et al. (2003), Zhang et al. (2008), Li et al. (2009), Cui et al. (2012) showed that increased forest cover in watersheds decreased water yield.

As both surface runoff and baseflow are the main components that contribute to water yield, we expected greater infiltration rate in ESAs scenario, for infiltration rate in forest areas is greater than in other land covers, e.g., sugarcane and pasture (Liu et al., 2012). Higher infiltration rate will increase baseflow, because in this scenario areas previously occupied with other land uses were now occupied with native vegetation. On the other hand, forest evapotranspiration will consume more water (PBIAS of evapotranspiration equal to +3.5%, Figure 14), because it is known that the forest is the surface with higher rates of evapotranspiration, higher than all the other vegetation types and also higher than a liquid's surface (Birkinshaw et al., 2011). Roots, especially of larger trees, increase water absorption from baseflow and, consequently, decrease water yield in the watershed, which may be seen in Figure 15, as the water content in the soil decreased in the studied period (-14.1%). Differently, with scenario change, this type of land cover provides greater resistance to runoff and, consequently, this component had a lower contribution to water yield in the watershed (-45.8%).

Table 5. PBIAS of hydrological variables analyzed between the two scenarios (current use and ESAs) in Pinhal River watershed, in the 2012-2014 period.

| Variable | Current use | ESAs scenario | PBIAS (%) |
|---|---|---|---|
| Discharge ($m^3/s$) | 119.1 | 105.3 | -11.6 |
| Surface runoff (mm) | 570.4 | 309.1 | -45.8 |
| Evapotranspiration (mm) | 1,993.2 | 2,062.3 | +3.5 |
| Soil water content (mm) | 8,279.8 | 7,113.5 | -14.1 |
| Water yield (mm) | 1,471.4 | 1,187.9 | -19.3 |

The influence of forest recovery in the hydrological regime can also be analyzed separately in two different periods. Comparing evapotranspiration demand independently in the wet period (October to March, Figure 14a) and dry period (April to September, Figure 14b), the difference between the two scenarios is even greater. In the wet period the difference is +1.3%, whereas in the dry period this difference is +8.2%. In the wet period available water in the soil (Figure 15a) compensates increased evapotranspiration demand of vegetation, even with increased forest cover (ESAs scenario), which contributes to lower water losses through evapotranspiration in the watershed (Figure 14a). In the dry period when SW is lower (Figure 15b), large forest vegetation access more easily underground water than a small vegetation, having, therefore, greater evapotranspiration demand and reducing water yield in the watershed. Based on results obtained from more than 90 experimental micro watersheds in different parts of the world, Bosch & Hewlett (1982) asserted that deforestation decreases evapotranspiration, which results in more water available in the soil and in streamflow. On the other hand, reforestation decreases streamflow at the watershed scale. It is worth mentioning, however, that these results vary from place to place and are often unpredictable (BROWN et al., 2005).

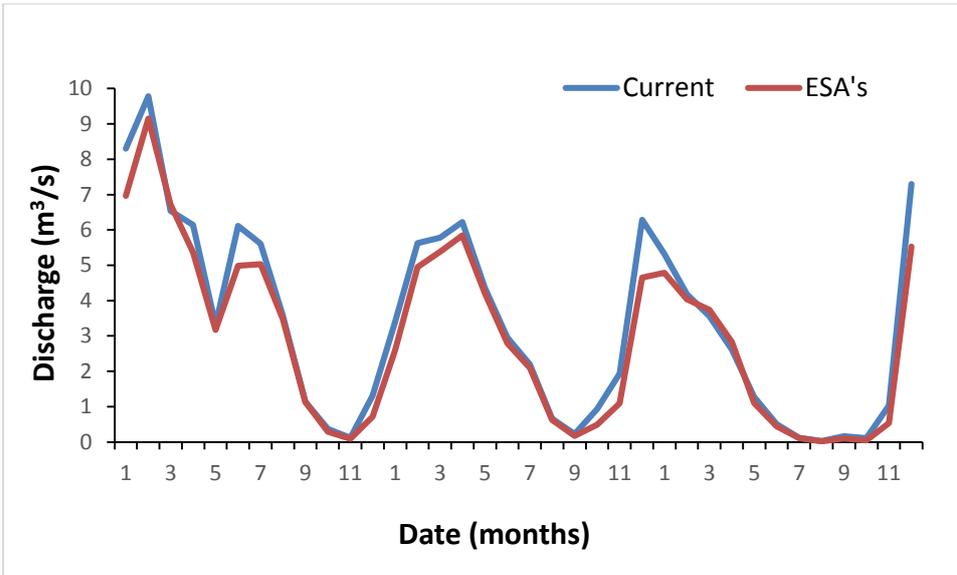

Figure 11 – Pinhal River watershed discharge comparison between the two scenarios.

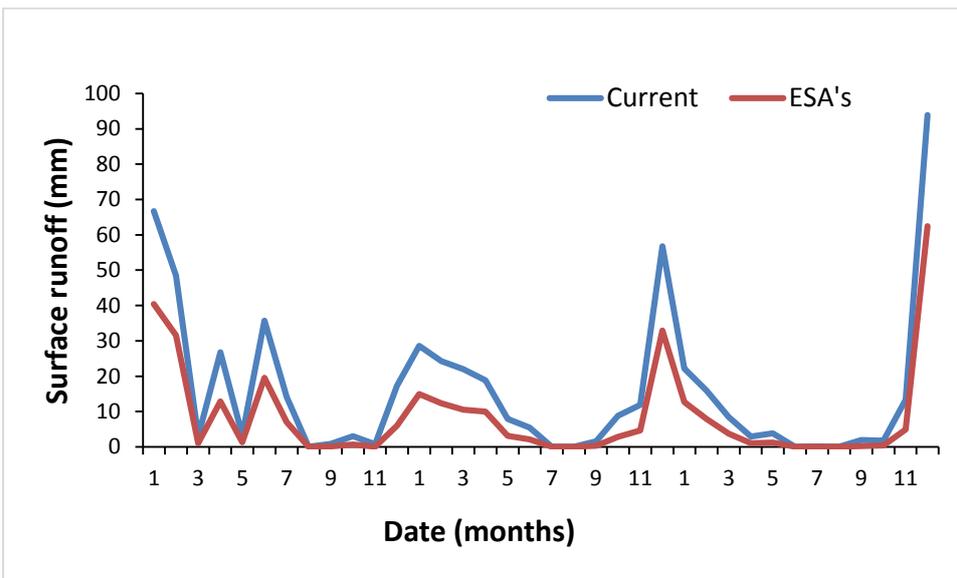

Figure 12 – Pinhal River watershed surface runoff in the two scenarios.

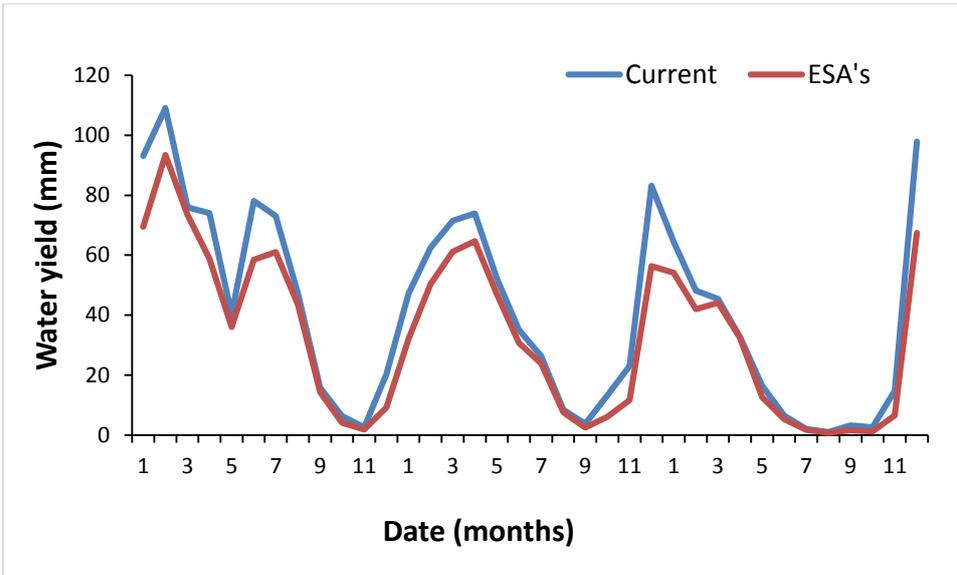

Figure 13 – Comparison of water produced in Pinhal River watershed between the two scenarios.

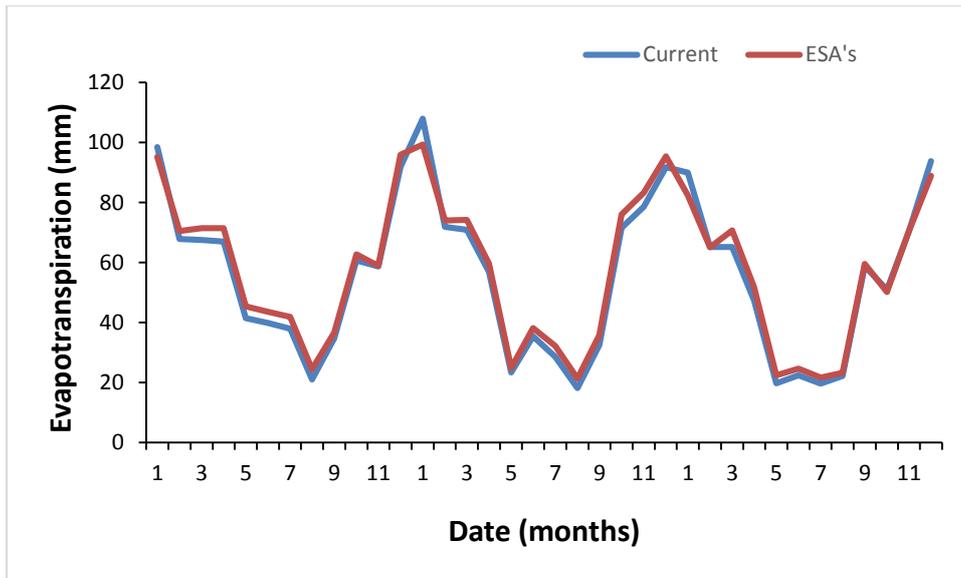

Figure 14 – Pinhal River watershed evapotranspiration in two different scenarios.

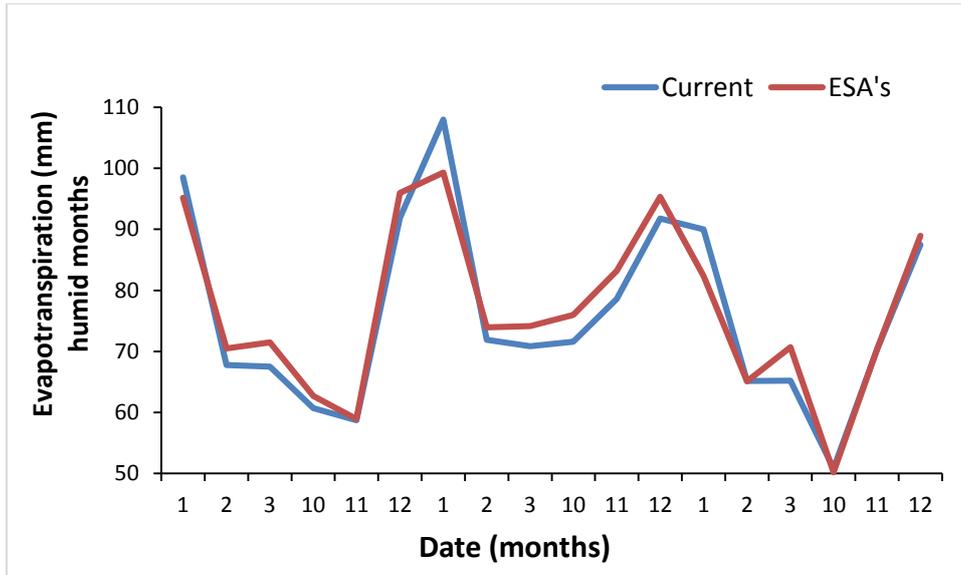

Figure 14a – Pinhal River watershed wet season evapotranspiration in two different scenarios [PBIAS = +1.3%].

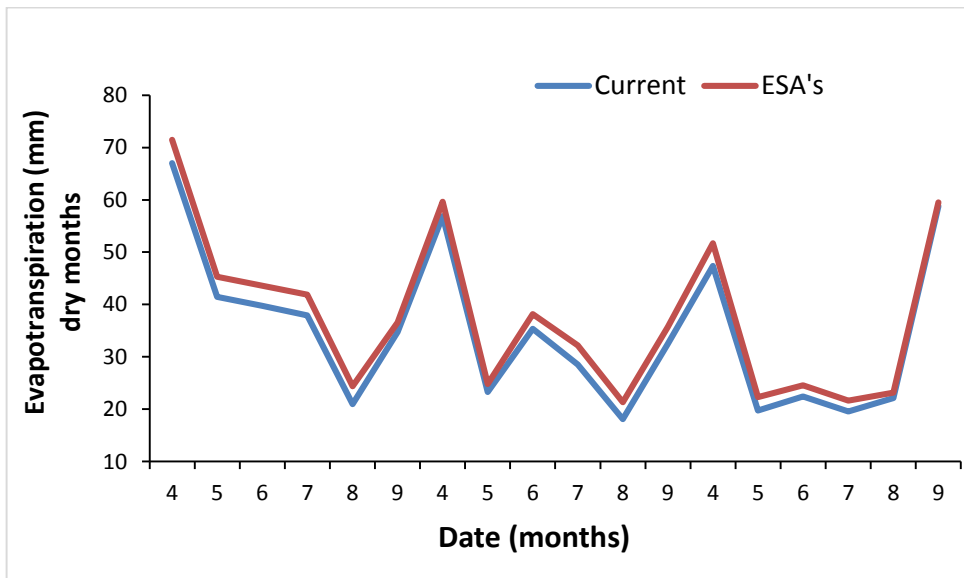

Figure 14b – Pinhal River watershed dry season evapotranspiration in two different scenarios [PBIAS = +8.2%].

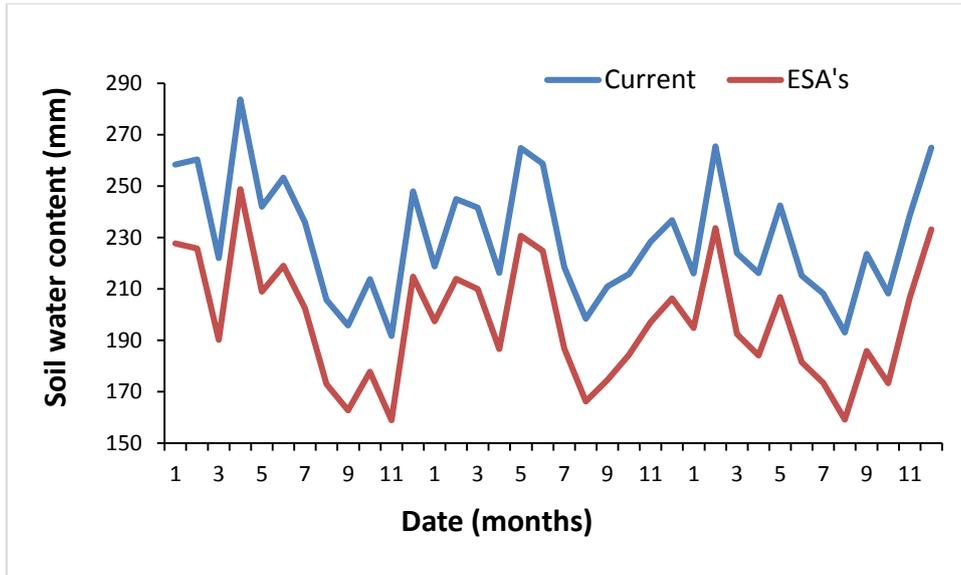

Figure 15 – Pinhal River watershed soil water content in two different scenarios.

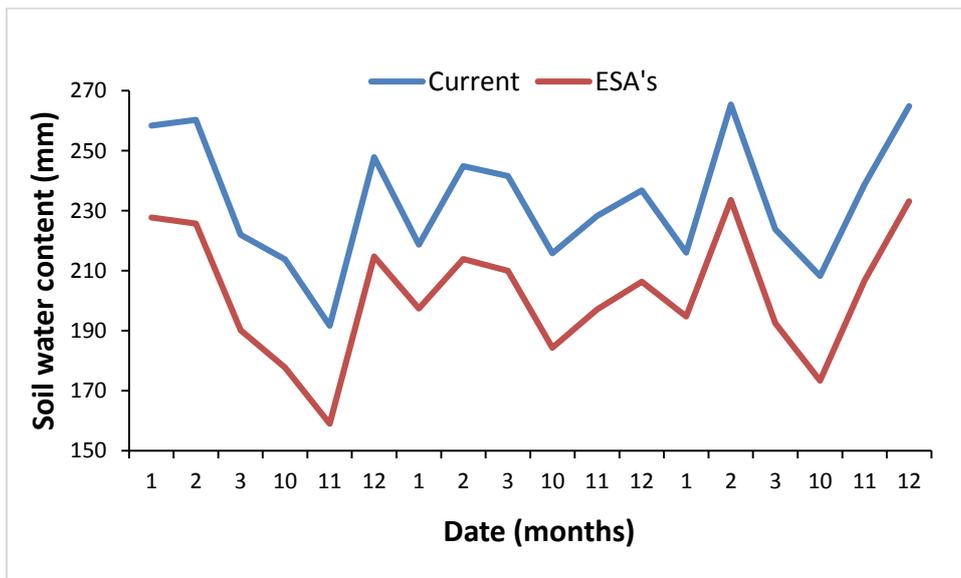

Figure 15a – Pinhal River watershed soil water content in two different scenarios [PBIAS = -13.3%].

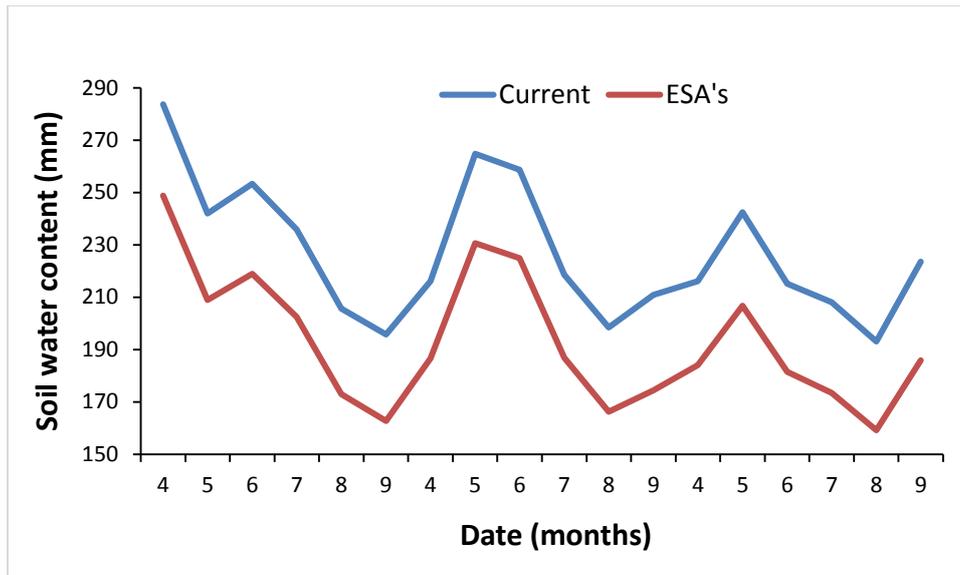

Figure 15b – Pinhal River watershed soil water content during dry season in two different scenarios [PBIAS = -14.9%].

    Figure 16 shows spatial distribution of the hydrological regime variation (surface runoff, evapotranspiration, soil water content and water yields) at sub-watersheds scale between scenarios. The influence of land-use change on the hydrological regime is more visible in some of the sub-watersheds than others at the watershed scale. These variations were smaller in upstream sub-watersheds and as with sediment yield, major variations occurred in sub-watersheds with greater forest cover when we compare current scenario with ESAs scenario. Watersheds hydrological regime is the result of complex interactions between climate (wet versus dry years), plant physiological properties (e.g., leaf area and successional stages) and soil type (ANDREASSIAN, 2004). According to Singh & Mishra (2012), these and other factors together make hydrological effects of forests a markedly different scenario.

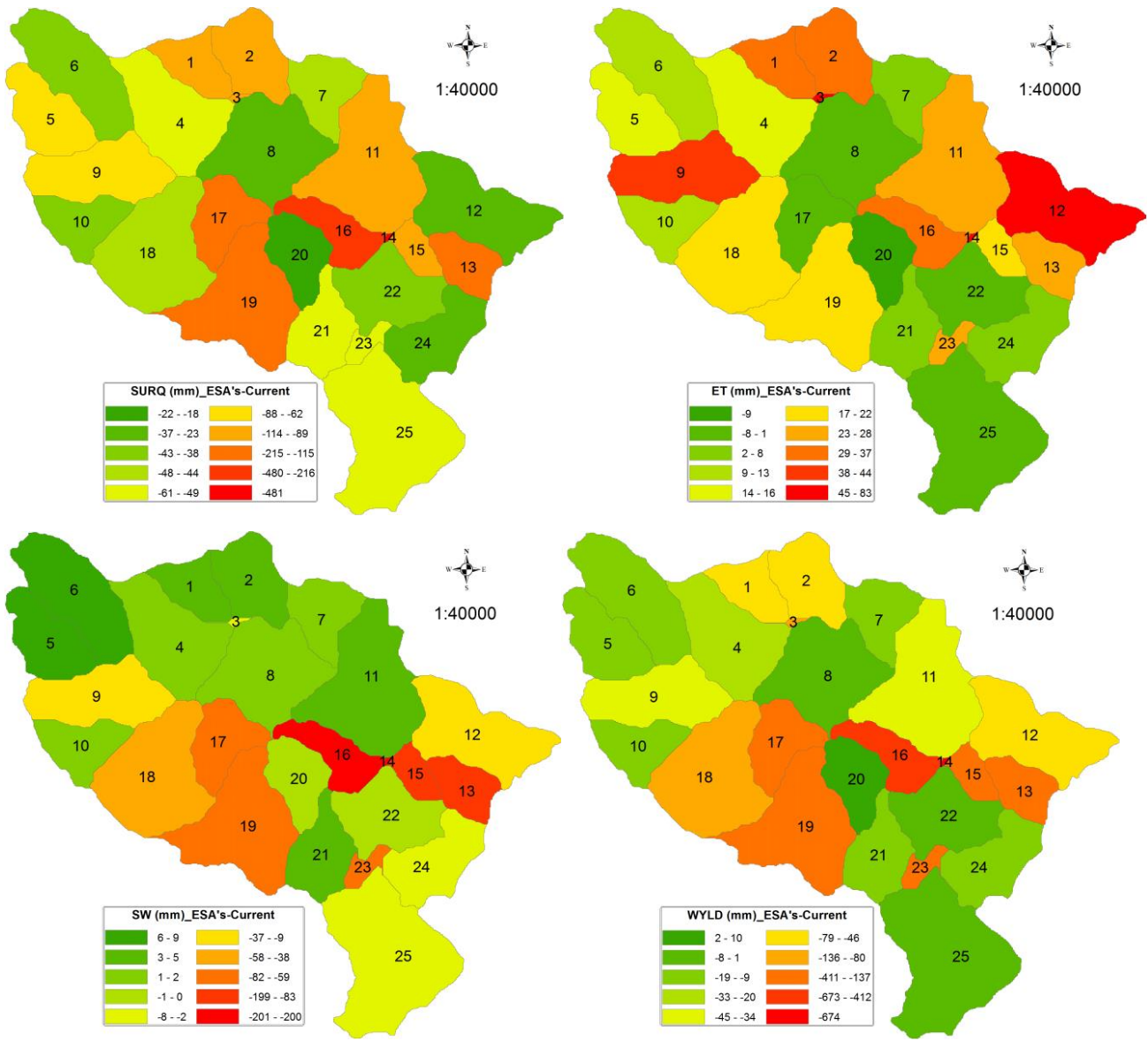

Figure 16. Spatial variations of average annual hydrological regime at sub-watershed scale between the two scenarios. SURQ (surface runoff - mm), ET (evapotranspiration - mm), SW (soil water content - mm), WYLD (water yield - mm).

## 4. CONCLUSION

The role of forests in watersheds hydrological cycle and water yield is controversial. Although reducing sediment yield as the results obtained from simulation of different scenarios show (PBIAS = -54%), for it offers the soil greater protection, its influence on increasing and maintaining streamflow is questionable, because the results obtained from this study also showed that increased forest cover decreased water yield in the watershed in -19.3% (PBIAS) due mostly to its greater evapotranspiration capacity (+3.5%), this demand being even greater during dry season (+8.2%). Simulation results lead us to conclude that the impacts of land-use change on hydrological processes are complex and their consequences are not equal in all situations and with the same intensity.

**Acknowledgments**

UNICAMP Espaço da Escrita project/General Coordination for the English translation of this article.

**Funding:** This work was funded by São Paulo Research Foundation (FAPESP) [1grant #2013/02971-3].

# REFERENCES

ABBASPOUR, K.C.; ROUHOLAHNEJAD, E.; VAGHEFI, S.; SRINIVASAN, R.; YANG, H.; KLØVE, B. A continental-scale hydrology and water quality model for Europe: Calibration and uncertainty of a high-resolution largescale SWAT model. **Journal of Hydrology**, v. 524, n. 5, p. 733-752, 2015.

ADAMI, S. F.; COELHO, R. M.; CHIBA, M. K.; MORAES, J. F. L. Environmental fragility and susceptibility mapping using geographic information systems: applications on Pinhal River watershed (Limeira, State of São Paulo). **Acta Scientiarum.** Technology (printed), v. 34, p. 433-440, 2012.

ANDREASSIAN, V., 2004. Waters and forests: from historical controversy to scientific debate. **Journal of Hydrology** 291, 1–27.

ARNOLD J.; WILLIAMS J.; MAIDMENT D. (1995) Continuous-time water and sediment-routing model for large basins. **J Hydraulic Eng.**, 121:171–183.

ARNOLD, J.G., SRINIVASAN, R., MUTTIAH, R.S., WILLIAMS, J.R., 1998. Large area hydrologic modelling and assessment part I: model development. J. Am. **Water Resour. Assoc.** 34 (1), 73–89.

ARNOLD, J.G., MORIASI, D.N., GASSMAN, P.W., ABBASPOUR, K.C., WHITE, M.J., SRINIVASAN, R., SANTHI, C., HARMEL, R.D., VAN GRIENSVEN, A., VAN LIEW, M.W., KANNAN, N., JHA, M.K., 2012. SWAT: model use, calibration, and validation. Trans. **ASABE** 55 (4), 1491–1508.

BACELLAR, L. de A, P., 2005. O papel da floresta no regime hidrológico de bacias hidrográficas. **Depto. de Geologia da UFOP** - Ouro Preto/MG: 1-39

BEST A., ZHANG, L., MCMAHOM T., WESTERN, A, VERTESSY R. 2003. A critical review of paired catchment studies with reference to seasonal flow and climatic variability. Australia, CSIRO Land and Water Technical. MDBC Publication, 56 p. (**Technical Report 25/03**).

BIRKINSHAW, S. J.; BATHURST, J. C.; IROUMÉ, A.; PALACIOS, H. The effect of forest cover on peak flow and sediment discharge—an integrated field and modelling study in central–southern Chile. **Hydrol. Process.** 25, 1284–1297 (2011).

BOSCH, J.M., HEWLETT, J.D., 1982. A review of catchment experiments to determine the effect of vegetation changes on water yield and evapotranspiration. **Journal of Hydrology** 55 (1/4), 3–23.

BROWN, A. E.; ZHANG B., L.; MCMAHONC, T. A. WESTERNC, A. W.; VERTESSYB, R. A. A review of paired catchment studies for determining changes in water yield resulting from alterations in vegetation. **Journal of Hydrology** 310 (2005) 28–61.

BRUIJNZEEL, L.A. (1990). **Hydrology of Moist Tropical Forests and Effects of Conversion**: a State of Knowledge Review. Humid Tropics Programme, IHP-UNESCO, Paris, and Vrije Universiteit, Amsterdam, 224 pp.

BRUIJNZEEL, L.A. Hydrological functions of tropical forests: not seeing the soil for the trees? **Agriculture, Ecosystems and Environment** 104 (2004) 185–228.

BUYTAERT, W.; CÉLLERI, R.; DE BIÈVRE, B., CISNEROS, F.; WYSEURE, G.; DECKERS, J.; HOFSTEDE, R. Human impact on the hydrology of the Andean páramos. **Earth-Science Reviews** 79 (2006) 53–72.

CHO, S. M.; JENNINGS, G. D.; STALLINGS, C.; DEVINE, H. A. GIS-basead water quality model calibration in the Delaware river basin. **ASAE**, St. Joseph, Michigan, 1995. (ASAE Microfiche, 952404)

CUI, X.; LIU, S.; WEI, X. Impacts of forest changes on hydrology: a case study of large watersheds in the upper reaches of Minjiang River watershed in China. **Hydrol. Earth Syst. Sci.**, 16, 4279–4290, 2012

DECHMI, F. & SKHIRI, A. Evaluation of best management practices under intensive irrigation using SWAT model. **Agricultural Water Management**, 2013, vol. 123, issue C, pages 55-64.

DOUGLAS, M. K.; SRINIVASAN, R.; ARNOLD, J. Soil and water assessment tool (swat) model: Current developments and applications. **T. Asabe**, 2010, 53, 1423–1431.

Emam, A. R.; Kappas, M.; Nguyen, L. H. K. and Renchin, T. Hydrological Modeling in an Ungauged Basin of Central Vietnam Using SWAT Model. Manuscript under review for **journal Hydrol. Earth Syst. Sci**. Published: 18 February 2016.

FUKUNAGA, D. C. et al. Application of the SWAT hydrologic model to a tropical watershed at Brazil. **Catena**, 125:206-213, 2015.

GOURLAY, D; SLEE, B. Public preferences for landscape features: a case study of two Scottish environmental sensitive areas. **Journal of Rural Studies**, v. 14, n.2, p. 249-263, 1998.

Gupta, H. V., Sorooshian, S., & Yapo, P. O. (1999). Status of automatic calibration for hydrologic models: Comparison with multilevel expert calibration. **Journal of Hydrologic Engineering**, 4(2), 135-143. DOI: 10.1061/(ASCE)1084-0699(1999)4:2(135)

HUANG, M.; ZHANG, L.; GALLICHAND, J. Runoff responses to afforestation in a watershed of the Loess Plateau, China. **Hydrol. Process.** 2003, 17, 2599–2609.

KUWAJIMA, J. I.; ARANTES, D. M.; ESTIGONI, M. V.; MAUAD, F. F. Proposta da aplicação do modelo SWAT como ferramenta complementar de gerenciamento de recursos hídricos e estimativa de assoreamento em reservatórios. In: **XIV World Water Congress**, 2011, Porto de Galinhas - PE. Adaptative Water Management: Looking to the future, 2011. p. 134-134.

KOCH, F. J.; VAN GRIENSVEN, A.; UHLENBROOK, S.; TEKLEAB, S.; TEFERI, E. The Effects of Land use Change on Hydrological Responses in the Choke Mountain Range (Ethiopia) - A new Approach Addressing Land Use Dynamics in the Model SWAT. **International Environmental Modelling and Software Society (iEMSs)** 2012 International Congress on Environmental Modelling and Software Managing Resources of a Limited Planet, Sixth Biennial Meeting, Leipzig, Germany, 2012.

LEINZ, V.; LEONARDOS, O. H. **Glossário geológico**. 2. ed. São Paulo: Companhia Editora Nacional, 1977. 236p.

LESSA, L. G. F.; SILVA, A. F.; ZIMBACK, C. R. L. MACHADO, R.E.; LIMA, S. L. Spatial distribution of sediment and water yield, applying geostatistical tools. **Journal of Environmental Science and Water Resources**, Vol. 3(3), p. 032 - 039, 2014.

LIMA, W. P., 2010. A Silvicultura e a Água: Ciência, Dogmas, Desafios. **Cadernos do Diálogo**, Vol. I. Instituto BioAtlântica, Rio de Janeiro. 64 p.

LIN, B.; CHEN, X.; YAO, H.; CHEN, Y.; LIU, M.; GAO, L.; JAMES, A. Analyses of landuse change impacts on catchment runoff using different time indicators based on SWAT model. **Ecol. Indic.**, 58 (2015), pp. 55–63.

LIU, Y., YANG, W., YU, Z., LUNG, I., & GHARABAGHI, B. (2015). Estimating sediment yield from upland and channel Erosion at a watershed scale using SWAT. **Water Resources Management**, Volume 29, Issue 5, pp 1399–1412.

LIU, Z.; LANG, N.; WANG, K. Infiltration Characteristics under Different Land Uses in Yuanmou Dry-Hot Valley Area. **In Proceedings of the 2nd International Conference on Green Communications and Networks 2012 (GCN 2012):** Volume 1; Yang, Y., Ma, M., Eds.; Springer: Berlin, Germany, 2013; Volume 223, pp. 567–572.

MACHADO, R. E. & VETTORAZZI, C. A. Sediment yield simulation for the Marins watershed, State of São Paulo, Brazil. **Rev. Bras. Ciênc. Solo** [online]. 2003, vol.27, n.4, pp.735-741. ISSN 1806-9657. http://dx.doi.org/10.1590/S0100-06832003000400018.

MACHADO, R. E.; VETTORAZZI, C. A.; CRUCIANI, D. E. Simulação de Escoamento em uma Microbacia Hidrográfica utilizando Técnicas de Modelagem e Geoprocessamento. **Revista Brasileira de Recursos Hídricos**, v. 8, n.1, p. 147-155, 2003.

MORIASI, D.N.; ARNOLD, J.G.; VAN LIEW, M.W.; BINGNER, R.L.; HARMEL, R.D.; VEITH, T.L. Model evaluation guidelines for systematic quantification of accuracy in watershed simulations. **Transactions of the ASABE**. v. 50, n. 3, p. 885-900, 2007.

NAGHETTINI, M.; PINTO, E.J.A. **Hidrologia Estatística**. Belo Horizonte: CPRM, 2007. 552 p.

NASH, J. E.; SUTCLIFFE, J. V. River flow forecasting through conceptual models: a discussion of principles. **Journal of Hydrology**, 10(3):282-290, 1970.

NDUBISI, F.; DE MEO T.; DITTO, N.D. Environmentally sensitive area: a template for developing greenway corridors. **Landscape Urban Planning**, v.33, p.159-177, 1995.

OLIVEIRA, J. B. **Solos da folha de Piracicaba**. Campinas, Instituto Agronômico, 1999. 173p. (Boletim Científico, 48).

PETERSON, J. R.; HAMLETT, J. M. Hydrologic calibration of the SWAT model in a watershed containing fragipan soils. **Journal of the American Water Resources Association**, v.34, n.3, p.531-544, 1998.

ROCHA, J.; ROEBELING, P.; RIAL-RIVAS, M.E. Assessing the impacts of sustainable agricultural practices for water quality improvements in the Vouga catchment (Portugal) using the SWAT model. **Sci. Total Environ.**, 536 (2015), pp. 48–58 http://dx.doi.org/10.1016/j.scitotenv.2015.07.038.

ROSENTHAL, W. D.; SRINIVASAN R.; ARNOLD, J. G. Alternative River Management Using a Linked GIG-Hydrology Model. **Transactions of the ASAE**, v.38, n.3, p.783-790, 1995.

RUBIO, J.L. Desertification: evolution of a concept. In: FANTECHI, R.; PETER, D.; BALABANIS, P.; RUBIO, J.L. (Eds.) EUR 15415 **Desertification in a European**

**Physical and Socio-economic Aspects**, Brussels, Luxembourg. Office Publications of the Europen Communities, p.5-13, 1995.

SINGH, S. & MISHRA, A. Spatiotemporal analysis of the effects of forest covers on water yield in the Western Ghats of peninsular India. **Journal of Hydrology**, 446–447 (2012) 24–34.

SRINIVASAN, R.; ARNOLD J. G. Integration of a basin-scale water quality model with GIS. **Water Resources Bulletin**, v.30, n.3, p.453-462, 1994.

SUN, G.; ZHOU, G.; ZHANG, Z.; WEI, X.; MCNULTY, S. G.; VOSE, J. Potential water yield reduction due to forestation across China. **Journal of Hydrology** (2006) 328, 548– 558.

TERAMOTO, E.R. Relações solo, substrato geológico e superfícies geomórficas na microbacia do ribeirão Marins. Piracicaba, 1995. 93p. Dissertação (Mestrado) – Escola Superior de Agricultura "Luiz de Queiroz", Universidade de São Paulo.

TUCCI, C. E. M. Flow regionalization in the upper Paraguay basin, Brasil. **Hydrological Sciences Journal**, v. 40, n.4, p. 485 -497, 1995.

TUCCI, C. E. M. **Regionalização de Vazões** – Porto Alegre. Ed. Universidade/UFRGS, 2002.

VAN LIEW, M. W. et al. Suitability of SWAT for the conservation effects assessment project: a comparison on USDA-ARS watersheds. **Journal of Hydrological Research**, 12(2):173- 189, 2007.

ZHANG, P.; LIU, R.; BAO, Y.; WANG, J.; YU, W.; SHEN, Z. Uncertainty of SWAT model at different DEM resolutions in a large mountainous watershed. **Water Res.**, 53 (2014), pp. 132-144 http://dx.doi.org/10.1016/j.watres.2014.01.018

Zhang, Y., Liu, S., Wei, X., Liu, J., and Zhang, G.: Potential impact of afforestation on water yield in the subalpine region of southwesternchina, **J. Am. Water Resour. Assoc.**, 44 (2008), pp. 1144–1153.